\documentclass[10pt]{article}
\usepackage{amsmath,amssymb,cite,epsfig}
\begin{document}

\title{\begin{flushright}
\small IISc/CHEP/01/10
\end{flushright}
\vspace{0.25cm} Thermal Correlation Functions of Twisted Quantum Fields}
\author{Prasad Basu\footnote{prasad@cts.iisc.ernet.in}, 
Rahul Srivastava\footnote{rahul@cts.iisc.ernet.in} 
and Sachindeo Vaidya\footnote{vaidya@cts.iisc.ernet.in}\\
\begin{small}Centre for High Energy Physics, Indian
Institute of Science,
Bangalore, 560012, India.
\end{small}}
\date{}
\maketitle

\begin{abstract}
We derive the thermal correlators for twisted quantum fields on noncommutative spacetime. We show 
that the thermal expectation value of the number operator is same as in commutative spacetime, 
but that higher correlators  are sensitive to the noncommutativity parameters $\theta^{\mu\nu}$.
\end{abstract}

\vspace{2mm}

General arguments involving classical gravity and quantum uncertainties suggest that spacetime
structure should be ``granular'' at very short distances \cite{dfr}. A specific model for this granularity 
is realized by the Groenewold-Moyal (GM) plane, where instead of the usual pointwise product 
$(f\cdot g)(x)$ on ${\mathbb R}^{d+1}$, one works with the noncommutative product
$(f\ast g)(x) = f(x) e^{\overleftarrow{\partial}_{\mu} \theta^{\mu\nu}  \overrightarrow{\partial}_{\nu} } g(x)$.
A particularly important feature of GM plane is that Poincar\'e symmetries are automorphisms of the spacetime, albeit with a twisted coproduct $ \Delta_{\theta}(\Lambda)$
 \cite{cha} instead of the usual one. This in turn leads to deformation between the canonical (anti-)commutation relations \cite{bal,sachin} in quantum field theory:  
\begin{eqnarray} \left.
\begin{array}{l l l}
a_{\bold{p}}a_{\bold{q}} &=& \eta e^{ip\wedge q}a_{\bold{q}}a_{\bold{p}}, \quad a^{\dagger}_{\bold{p}}
a^{\dagger}_{\bold{q}} = \eta e^{ip\wedge q}a^{\dagger}_{\bold{q}}a^{\dagger}_{\bold{p}} \\
a_{\bold{p}}a^{\dagger}_{\bold{q}} &=& \eta e^{-ip\wedge q}a^{\dagger}_{\bold{q}}a_{\bold{p}} + 
(2\pi)^{3}2p_{0} \delta^{3}(\bold{p}-\bold{q}) 
\end{array}\right\}
\begin{array}{l l l}
\qquad \text{where} \quad p^{\mu} = (p^{0}, \bold{p}), \quad p\wedge q = p_\mu \theta^{\mu\nu}q_\nu, \\
\qquad \text{and} \quad \eta = \pm 1  \quad \text{for bosons/fermions}.
\end{array}
\label{tcom}
\end{eqnarray} 
Since effects of noncommutativity become important at high energies, we expect that there may be  
important implications in early cosmology with its attendant high temperatures. To this end, it is 
important to formulate the thermodynamics of such quantum field theories. Noncommutative spacetimes 
contribute an additional subtlety to this issue, in that the usual facility of working with a finite volume $V$ 
and then taking $V\rightarrow \infty$ is not available to us. Thus the appropriate starting point for 
any discussion of quantum thermodynamics is the KMS condition (see for instance, \cite{kms,haag}). 
We shall demonstrate two different (but equivalent) methods of computing thermal correlators, based 
on dual versions of the KMS condition.

Given an operator $A$ (which may for instance be constructed from products of quantum fields, or 
from products of creation or annihilation operators) in the Heisenberg representation, its time evolution 
is given by $A(\tau) = e^{i \mathcal{H} \tau} A  e^{-i \mathcal{H} \tau}, \quad \mathcal{H} = H -\mu N$
where $\mathcal{H}$ is the grand canonical Hamiltonian. It is important to emphasize that the 
$\tau$ appearing in the above equation is not the coordinate time $x^{0}$, but the parameter of 
time evolution \cite{bgmt}.

For any two operators $A$ and $B$, we can define the retarded function
\begin{equation} 
G_{AB}(\tau-\tau') \equiv -i\theta(\tau-\tau')\langle\langle  [A(\tau), B(\tau')]\rangle\rangle  
= -i\theta (\tau-\tau')[\langle A(\tau)B(\tau')\rangle -\eta' \langle B(\tau')A(\tau)\rangle ]
\label{green}
\end{equation}
where $\theta(x)$ is the Heavyside step-function. In situations where the Gibbs state $\rho$ can 
be defined (for example for systems in a finite volume $V$), the thermal average of any operator $X$ is 
$\langle  X  \rangle = \frac{{\rm Tr}\, [e^{-\beta\mathcal{H}}X]}{Z} \equiv {\rm Tr}\, [\rho  X]$, where 
 $\rho = \frac{e^{-\beta\mathcal{H}}}{Z}$ and $Z = {\rm Tr}\,[e^{-\beta\mathcal{H}}].$
 Advanced and causal functions can be defined similarly \cite{zubarev}. 

We will instead make use of the relation between 
$G_{AB}(\tau-\tau')$, the thermal correlation functions $\mathcal{F}_{AB}(\tau-\tau')=
\langle A(\tau)B(\tau')\rangle$ and $\mathcal{F}_{BA}(\tau-\tau')=\langle B(\tau')A(\tau)\rangle$, and the spectral density $J_{BA}(\omega)$ defined by 
\begin{equation}
\mathcal{F}_{BA}(\tau-\tau') = \int^\infty _{-\infty} J_{BA}(\omega)e^{-i\omega(\tau-\tau')} 
 d\omega.
 \label{spec1}
\end{equation}
Thermodynamic equilibrium (i.e. objects like $G_{AB}$ and $\mathcal{F}_{AB}$ are functions of 
$(\tau-\tau')$ only) and cyclicity of trace imply that
\begin{equation}
\mathcal{F}_{AB}(\tau-\tau') = \int^\infty _{-\infty} J_{BA}(\omega)e^{\beta\omega}
e^{-i\omega(\tau-\tau')}d\omega, 
\label{spec2}
\end{equation}
i.e. $\mathcal{F}_{AB}$ and $\mathcal{F}_{BA}$ satisfy the Fourier space version of the KMS 
condition \cite{haag}. We will use this as our starting point, rather than assume the existence of the 
Gibbs state $\rho$, thus circumventing the formal necessity of putting the system in a box of finite volume.

For evaluating correlators of interest, we will follow the strategy outlined in \cite{zubarev}. 
The $\tau$-independent function $\eta'$ shall be chosen so that $G_{AB}$ satisfies a 
conveniently simple differential equation, as we shall show below.

Heisenberg equations of motion for $A(\tau)$ and $B(\tau)$ imply that $G_{AB}$ satisfies
\begin{equation}
i\frac{dG_{AB}}{d\tau} = \delta(\tau-\tau') \langle A(\tau) B (\tau) - \eta' B(\tau)  A(\tau) ]\rangle
 + \langle\langle \{ A(\tau) \mathcal{H} - \mathcal{H} A(\tau); B(\tau')\} \rangle\rangle.
\label{motion}
\end{equation} 
The Fourier transform  
$G_{AB}(E) \equiv \frac{1}{2\pi} \int ^{\infty}_{-\infty}G_{AB}(\tau)e^{iE\tau}d\tau$ can be written as
\begin{equation}
G_{AB}(E) = \frac{1}{2\pi} \int ^{\infty}_{-\infty} J_{BA}(\omega)(e^{\beta\omega} - \eta') 
\frac{d\omega}{E-\omega +i\epsilon}
\label{fougre}
\end{equation}
using the integral representation $\theta(\tau-\tau') = \frac{i}{2\pi}\int ^{\infty}_{-\infty} 
\frac{e^{-ix(\tau-\tau')}}{x+i\epsilon}$ in (\ref{spec2}).

Using (\ref{fougre}) and the delta function representation $\delta(x) = \frac{1}{2\pi i} \left \{ \frac{1}{x-i\epsilon} -  \frac{1}{x+i\epsilon} \right\} $ we get
\begin{equation}
 G_{AB}(\omega + i\epsilon) - G_{AB}(\omega - i\epsilon) = -i J_{BA}(\omega)(e^{\beta\omega}-\eta'),
\label{specngre} 
\end{equation}
which in turn gives 
\begin{eqnarray}
\mathcal{F}_{BA}(\tau-\tau') &=& \int ^{\infty}_{-\infty} \frac{G_{AB}(E + i\epsilon)- G_{AB}(E - i\epsilon)}{e^{\beta E}-\eta ^{'}} e^{-iE(\tau-\tau')}dE, \label{corrba}\\
\mathcal{F}_{AB}(\tau' - \tau) &=& \int ^{\infty}_{-\infty} \frac{G_{AB}(E + i\epsilon)- G_{AB}(E - i\epsilon)}{e^{\beta E}-\eta ^{'}} e^{\beta E} e^{-i E(\tau-\tau')}dE.
\label{corrab}
\end{eqnarray}
For a perfect quantum gas, the (grand canonical) Hamiltonian  is
\begin{equation}
\mathcal{H}= H - \mu N = \int \frac{d^{3}\bold{k}}{(2\pi)^{3}2\omega _{\bold{k}}}(\omega _{\bold{k}} - \mu)a^{\dagger}_{\bold{k}}a_{\bold{k}} 
\label{haml} 
\end{equation}
where the $a^{\dagger}_{\bold{k}} $  and $a_{\bold{k}} $ satisfy (\ref{tcom}).
Substituting 
$A(\tau)=a_{\bold{p_{1}}}(\tau), B(\tau')=a^{\dagger}_{\bold{p_{2}}}(\tau')$ in (\ref{green}), we 
find that
\begin{equation}
G_{\bold{p_{1}}\bold{p_{2}}} \equiv  - i \theta (\tau -\tau')[\langle a_{\bold{p_{1}}}(\tau)
a^{\dagger}_{\bold{p_{2}}}(\tau')\rangle - \eta' \langle a^{\dagger}_{\bold{p_{2}}}(\tau')
a_{\bold{p_{1}}}(\tau)\rangle]
\label{twogreen}
\end{equation}
satisfies
\begin{equation}
i\frac{dG_{\bold{p_{1}}\bold{p_{2}}}}{d\tau} = (2\pi)^{3}2(p_{10})\delta(\tau -\tau')
\delta^{3}(\bold{p_{1}}-\bold{p_{2}}) + \left(\omega _{\bold{p_{1}}} - \mu\right) 
G_{\bold{p_{1}}\bold{p_{2}}}(\tau-\tau')
\label{twomotion}
\end{equation}
if we choose $\eta' = \eta  e^{-ip_{1}\wedge p_{2}} $.

The Fourier transform $ G_{\bold{p_{1}}\bold{p_{2}}}(E) $  
of $G_{\bold{p_{1}}\bold{p_{2}}} (\tau-\tau')$ is easily obtained:
\begin{equation}
G_{\bold{p_{1}}\bold{p_{2}}}(E) = \frac{1}{2\pi}\frac{(2\pi)^{3} 2(p_{10}) \delta^{3}(\bold{p_{1}} - 
\bold{p_{2}})}{E-\left(\omega _{\bold{p_{1}}}-\mu\right)}.
\label{gfinal}
\end{equation}

Using (\ref{corrba}) and putting $ \tau = \tau'$, we get 
\begin{equation}
\langle a^{\dagger}_{\bold{p_{2}}}a_{\bold{p_{1}}}\rangle = \frac{(2\pi)^{3} 2(p_{10}) 
\delta^{3}(\bold{p_{1}}-\bold{p_{2}})}{e^{\beta \left(\omega _{\bold{p_{1}}}-\mu\right)}-\eta e^{-ip_{1}\wedge p_{2}}}.
\end{equation}
Since $ p_{1}\wedge p_{2} = 0 $ if $ \bold{p_{1}} =  \bold{p_{2}} $, we have
\begin{equation}
\langle a^{\dagger}_{\bold{p_{2}}}a_{\bold{p_{1}}}\rangle = \frac{(2\pi)^{3} 2(p_{10}) 
\delta^{3}(\bold{p_{1}}-\bold{p_{2}})}{e^{\beta \left(\omega _{\bold{p_{1}}}-\mu\right)}-\eta},
\label{twopoint}
\end{equation}
which is same as the commutative correlation function.

This result is not unexpected: translational invariance 
forces this upon us. Higher correlators however will not be so severely restricted by 
translational invariance. For instance, to calculate $\langle a^{\dagger}_{\bold{p_{1}}} 
a^{\dagger}_{\bold{p_{2}}} a_{\bold{p_{3}}} a_{\bold{p_{4}}}\rangle$, we substitute 
$A(\tau)= a_{\bold{p_{4}}}(\tau)$ and $B(\tau')= a^{\dagger}_{\bold{p_{1}}}(\tau') 
a^{\dagger}_{\bold{p_{2}}}(\tau')a_{\bold{p_{3}}}(\tau') $ in (\ref{green}):
\begin{equation}
G_{\bold{p_{4}}\bold{p_{1}}\bold{p_{2}}\bold{p_{3}}} = -i \theta(\tau - \tau')\left[\langle 
a_{\bold{p_{4}}}(\tau)a^{\dagger}_{\bold{p_{1}}}(\tau')a^{\dagger}_{\bold{p_{2}}}(\tau') 
a_{\bold{p_{3}}}(\tau')\rangle  - \eta' \langle a^{\dagger}_{\bold{p_{1}}}(\tau') 
a^{\dagger}_{\bold{p_{2}}}(\tau')a_{\bold{p_{3}}}(\tau')a_{\bold{p_{4}}}(\tau) \rangle\right].
\label{4green}
\end{equation}
This satisfies
\begin{eqnarray} 
i \frac{dG_{\bold{p_{4}}\bold{p_{1}}\bold{p_{2}}\bold{p_{3}}}}{d\tau} & = & \delta (\tau -\tau') 
(2\pi)^{3}\left[2(p_{10})\delta ^{3}(\bold{p_{4}}-\bold{p_{1}}) \langle a^{\dagger}_{\bold{p_{2}}}(\tau) 
a_{\bold{p_{3}}}(\tau) \rangle \right. \nonumber \\
& + & \left.   2 \eta (p_{20}) \delta ^{3}(\bold{p_{4}}-\bold{p_{2}}) 
e^{-ip_{4}\wedge p_{1}} \langle a^{\dagger}_{\bold{p_{1}}}(\tau)a_{\bold{p_{3}}}(\tau)\rangle \right] 
+ \left(\omega _{\bold{p_{4}}}-\mu\right)G_{\bold{p_{4}}\bold{p_{1}}\bold{p_{2}}\bold{p_{3}}}
\label{fourgreen}
\end{eqnarray}
for the choice $ \eta'= \eta e^{-ip_{4}\wedge (p_{1}+ p_{2} - p_{3})}$. \\
The Fourier transform $ G_{\bold{p_{4}}\bold{p_{1}}\bold{p_{2}}\bold{p_{3}}}$ is
\begin{eqnarray} 
G_{\bold{p_{4}}\bold{p_{1}}\bold{p_{2}}\bold{p_{3}}}(E) & = & \frac{1}{2\pi} \frac{(2\pi)^{3}}{E - \omega _{\bold{p_{4}}}}[2(p_{10})\delta ^{3}(\bold{p_{1}}-\bold{p_{4}})
\langle a^{\dagger}_{\bold{p_{2}}}(\tau)a_{\bold{p_{3}}}(\tau) \rangle  \nonumber \\ 
& + & 2 \eta (p_{20}) e^{i p_{1}\wedge p_{4}}\delta ^{3}(\bold{p_{2}}-\bold{p_{4}})\langle a^{\dagger}_{\bold{p_{1}}}(\tau)a_{\bold{p_{3}}}(\tau)\rangle].
\label{fougreen}
\end{eqnarray}
Using (\ref{corrba}) and putting $ \tau = \tau'$ we get 
\begin{eqnarray}
\langle a^{\dagger}_{\bold{p_{1}}}a^{\dagger}_{\bold{p_{2}}}a_{\bold{p_{3}}}a_{\bold{p_{4}}} \rangle 
& = & \frac{(2\pi)^{3}(2p_{10})}{[e^{\beta \left(\omega _{\bold{p_{1}}}-\mu\right)} - \eta]} 
\frac{(2\pi)^{3}(2p_{20})}{[e^{\beta \left(\omega _{\bold{p_{2}}}-\mu\right)} - \eta]} \left[\delta ^{3}
(\bold{p_{1}}-\bold{p_{4}})\delta ^{3}(\bold{p_{2}}-\bold{p_{3}}) \right.\nonumber \\ 
& + & \left. \eta e^{i p_{1}\wedge p_{2}}\delta ^{3}(\bold{p_{1}}-\bold{p_{3}})\delta ^{3}(\bold{p_{2}} - 
\bold{p_{4}}) \right].
\label{fin4cor}
\end{eqnarray}
This four-point correlator
differs from its commutative counterpart by appearance of the $\theta$-dependent phase 
$e^{i p_{1}\wedge p_{2}}$ in the second term, and leads to interesting changes in 
observables like Hanbury-Brown and Twiss correlations \cite{hbt}.

Higher correlators may also be calculated by similar techniques as above. The computations are 
tedious but straightforward. Alternately, one can evaluate them by using the direct (as opposed to Fourier) 
formulation of the KMS condition \cite{haag}. Let $\omega_{\beta,\mu}$ be a positive, 
linear, normalized map from the algebra of operators to ${\mathbb C}$.
For any two operators $A$ and $B$  we define two functions 
${\mathcal F}_{AB}^{\beta, \mu}(\tau)$ and ${\mathcal G}_{AB}^{\beta, \mu}(\tau)$ as 
\begin{eqnarray}
{\mathcal F}_{AB}^{\beta, \mu}(\tau)=\omega_{\beta, \mu}(B A(\tau)) - 
\omega_{\beta, \mu}(A)\omega_{\beta, \mu}(B) \nonumber \\
{\mathcal G}_{AB}^{\beta, \mu}(\tau)=\omega_{\beta, \mu}(A(\tau) B ) - 
\omega_{\beta, \mu}(A)\omega_{\beta, \mu}(B),
\end{eqnarray}  
where, $A(\tau)=e^{i{\cal H}\tau} A e^{-i{\cal H}\tau}$. The map 
$\omega_{\beta, \mu}$ is 
a thermal state corresponding to the inverse temperature $\beta$ and chemical potential $\mu$ if 
\begin{eqnarray}
{\mathcal G}_{AB}^{\beta, \mu}(\tau)= {\mathcal F}_{AB}^{\beta, \mu}(\tau+i\beta). 
\end{eqnarray} 
Consider the operators $a^{\#}_{\bold{p}_{i}} $ and 
$a^{\#}_{\bold{p}_{j}}$, which stand for either creation or annihilation operators corresponding to 
the momentum state $\bold{p}_{i} $ and $ \bold{p}_{j}$ respectively. We define their {\it twisted 
commutator} as
\begin{equation}
[a^{\#}_{\bold{p}_{i}},a^{\#}_{\bold{p}_{j}}]_{\theta} \equiv a^{\#}_{\bold{p}_{i}} a^{\#}_{\bold{p}_{j}} - 
\eta e^{i(\alpha_{\bold{p}_{i}\bold{p}_{j}}) p_{i}\wedge p_{j}} a^{\#}_{\bold{p}_{j}} a^{\#}_{\bold{p}_{i}} ,
\label{tscm}
\end{equation}
where $\alpha_{\bold{p}_{i}\bold{p}_{j}}$ is 1 if $a^{\#}_{\bold{p}_{i} } $ and $a^{\#}_{\bold{p}_{j}} $ are 
of same type ({\it i.e.} are both creation or both annihilation operators), else is equal to $-1$. 
The commutation relations (\ref{tcom}) imply that
\begin{eqnarray}
[a^{\#}_{\bold{p}_{i}},a^{\#}_{\bold{p}_{j}}]_{\theta} & = & 0  \quad \quad
\mbox{if $ a^{\#}_{\bold{p}_{i}} $ and $a^{\#}_{\bold{p}_{j}} $ are of same type} \nonumber \\  
 & = & - \eta (2\pi)^{3}2(p_{10}) \delta^{3}({\bold p_i} -{\bold p_j}) \quad  \mbox{if $a^{\#}_{\bold{p}_{i}}=a^{\dagger}_{\bold{p}_{i}}$ 
and $a^{\#}_{\bold{p}_{j}}=a_{\bold{p}_{j}}$} \nonumber\\
&=& (2\pi)^{3}2(p_{10})\delta^{3}({\bold p_i} -{\bold p_j})  \quad \mbox{if $a^{\#}_{\bold{p}_{i}}=a_{\bold{p}_{i}}$ 
and $a^{\#}_{\bold{p}_{j}}=a^{\dagger}_{\bold{p}_{j}}$} .
\end{eqnarray}
Using (\ref{twopoint}), we see that
\begin{equation}
\omega_{\beta, \mu}(a^{\#}_{\bold{p}_{i}} a^{\#}_{\bold{p}_{j}}) = 
\frac{[a^{\#}_{\bold{p}_{i}},a^{\#}_{\bold{p}_{j}}]_{\theta} }
{1-\eta e^{\alpha\beta \left(\omega _{\bold{p_{i}}}-\mu\right)}},
\label{EQ36}
\end{equation}
where
\begin{eqnarray}
\alpha&=&1 \quad \mbox{if $a^{\#}_{\bold{p}_{i}} $ is a creation operator} \nonumber\\
&=&-1 \quad \mbox{if $a^{\#}_{\bold{p}_{i}}$ is an annihilation operator}    .
\end{eqnarray}
 
To compute the $N$-point correlator $\omega_{\beta, \mu} (a^{\#}_{\bold{p}_{1}} 
a^{\#}_{\bold{p}_{2}} \cdots a^{\#}_{\bold{p}_{N}})$, we use (\ref{tscm}) repeatedly to bring 
$a^{\#}_{\bold{p}_{1}}$ to the right side of the sequence. Linearity of $\omega_{\beta,\mu}$ then gives us 
\begin{eqnarray}
\omega_{\beta, \mu}(a^{\#}_{\bold{p}_{1}} a^{\#}_{\bold{p}_{2}}... a^{\#}_{\bold{p}_{N}}) &=&
\sum_{j=1}^{N-1}\eta^{j-1}e^{i\phi_j} [a^{\#}_{\bold{p}_{1}},a^{\#}_{\bold{p}_{j+1}}]_{\theta}\,
\omega_{\beta,\mu}(\widehat{a^{\#}_{\bold{p}_{1}}} a^{\#}_{\bold{p}_{2}} \cdots 
\widehat {a^{\#}_{\bold{p}_{j+1}}} \cdots a^{\#}_{\bold{p}_{N}}) \nonumber\\
&+&\eta^{N-1}e^{i\phi_{N}} \omega_{\beta,\mu}(a^{\#}_{\bold{p}_{2}}a^{\#}_{\bold{p}_{3}}\cdots a^{\#}_{\bold{p}_{N}}a^{\#}_{\bold{p}_{1}}) 
\end{eqnarray}
where \hspace{.5mm} $\widehat{}$ \hspace{.5mm} on an operator denotes the absence of this operator from the sequence. The phase 
$\phi_j$ is given by
\begin{equation}
\phi_j=\sum_{i=1}^{j}\alpha_{1i} \bold{p}_{1} \wedge \bold{p}_{i}.
\end{equation}
For $\tau=0$ the KMS condition implies that
\begin{equation}
\omega_{\beta, \mu}(A B)=\omega_{\beta, \mu}(B A(i\beta))
=\omega_{\beta, \mu}(B e^{-\beta{\cal H}}A e^{\beta{\cal H}} ).
\end{equation}
For the Hamiltonian (\ref{haml}),
using (\ref{EQ36}) we can finally write
\begin{equation}
\omega_{\beta, \mu}(a^{\#}_{\bold{p}_{1}} a^{\#}_{\bold{p}_{2}}... a^{\#}_{\bold{p}_{N}}) = 
\Bigl(\sum_{j=1}^{N-1}\eta^{j-1}e^{i\phi_j}
 \omega_{\beta,\mu} (a^{\#}_{\bold{p}_{1}}a^{\#}_{\bold{p}_{j+1}})
\omega_{\beta,\mu}( \widehat{a^{\#}_{\bold{p}_{1}}} a^{\#}_{\bold{p}_{2}} \cdots \widehat {a^{\#}_{\bold{p}_{j+1}}} \cdots a^{\#}_{\bold{p}_{N}}) \Bigr)\xi(\beta, N,\omega_{\bold p_1}),
\end{equation}
where $\xi(\beta, N,\omega_{\bold p_i})$ is given by
\begin{equation}
\xi(\beta, N,\omega_{\bold p_i})=\frac{1-\eta e^{\alpha\beta \left(\omega _{\bold{p_{i}}}-\mu\right)}}{1-\eta^{N-1}e^{i\phi_N}
e^{\alpha\beta \left(\omega _{\bold{p_{i}}}-\mu\right)}}.
\end{equation}
This is the thermal version of Wick's theorem adapted to twisted quantum fields: the $N$-point correlator
is expressed in terms of the $(N-2)$-point correlators. 

{\bf Acknowledgments:} We are grateful to Jayanta Bhattacharjee for pointing \cite{zubarev} to us. The work of P.B. is supported by a D.S.T grant.

\end{document}